\documentclass[twocolumn]{Articles_template_PIC_SRB}  
\shorttitle{Evidence of wave decay into $\mathcal Z$-mode electromagnetic radiation}
\shortauthors{Polanco-Rodr\'iguez, Krafft \& Savoini}

\usepackage{amsmath}
\usepackage{color}
\usepackage{float}
\usepackage{bm}
\usepackage{soul}
\usepackage{xcolor}
\usepackage{lipsum}
\usepackage{orcidlink}
\usepackage{enumitem}

\begin{document}
\graphicspath{{./Figures/}}







\title{Evidence of Langmuir/$\mathcal{Z}$-mode Wave Decay into $\mathcal{Z}$-mode Electromagnetic Radiation in the Solar Wind}

\author[1]{F. J. Polanco-Rodr\'{i}guez\orcidlink{0009-0005-2951-697X}}
\author[1,2]{C. Krafft\orcidlink{0000-0002-8595-4772}}
\author[1]{P. Savoini\orcidlink{0000-0002-2117-3803}}

\affil[1]{Laboratoire de Physique des Plasmas (LPP), CNRS, Sorbonne Université, 
Observatoire de Paris, Université Paris-Saclay, Ecole polytechnique, Institut Polytechnique de Paris, 91120 Palaiseau, France}
\affil[2]{Institut Universitaire de France (IUF)}

\date{January 2026}
\begin{abstract}
The nonlinear decay of Langmuir/$\mathcal{Z}$-mode waves  into electromagnetic $\mathcal{Z}$-mode wave radiation at the plasma frequency  is observed for the first time in the solar wind. This  finding was enabled by the unprecedented high-resolution electric and magnetic field measurements provided by the Radio Plasma Waves (RPW) instrument aboard the Solar Orbiter spacecraft, which encountered an electron beam associated with a Type III radio burst. The decay process is definitively identified through multiple lines of evidence: satisfaction of frequency and wavevector resonance conditions, strong phase coherence and temporal coincidence between the interacting waves, exclusion of competing mechanisms, and full agreement with theoretical predictions. Particle-in-cell simulations, conducted under close beam-plasma conditions, successfully reproduce the key features of the observations.  Notably, they suggest that the wave packet observed by Solar Orbiter may be trapped within an extended, nearly flat-bottomed density well, where the decay process is not overcome by wave scattering on random density fluctuations and subsequent mode conversion effects.

\end{abstract}

\section{Introduction}

Despite decades of routine observations of type III solar radio bursts in the solar wind, key questions regarding their radio emission mechanisms remain unanswered. With the advent of recent missions such as Solar Orbiter (\cite{Muller2020}) and Parker Solar Probe (\cite{Fox2016}), long-duration and well-resolved electric and magnetic waveforms are now recorded (\cite{Bale2016}, \cite{Maksimovic2020}). More specifically, recent advances in magnetometry have enabled the analysis of  highly resolved magnetic waveforms (\cite{Jannet2021}). Such measurements should help evidence and untangle the wave processes leading to electromagnetic waves radiated at the plasma frequency $f_p$ and its harmonic $2f_p$ during radio bursts.

Although Langmuir/$\mathcal{Z}$-mode waves (hereafter referred to as $\mathcal{LZ}$ waves) have been observed and studied in the solar wind using satellites such as Stereo and Wind (e.g., \cite{Bale1998, Bale2000}, \cite{Henri2009}, \cite{Malaspina2011}, \cite{GrahamCairns2013b}, \cite{Kellogg2013}), evidence for and detailed analyses of their magnetic signatures in the solar wind have emerged only recently.  In this context,  \cite{Larosa2022} used data from Parker Solar Probe to provide the first unambiguous observation of the magnetic component of a $\mathcal{Z}$-mode wave in the solar wind, which corresponds to the slow extraordinary electromagnetic mode with frequency below the local plasma frequency (e.g. \cite{Krauss-Varban1989}). The authors proposed the scattering of $\mathcal{LZ}$ waves on density fluctuations as the primary mechanism generating such waves.  More recently, \cite{Formanek2025} examined a high-amplitude wave packet observed by Solar Orbiter on 22 September 2022 at 13:52:28 UT. By analyzing electron beam velocity distribution functions (VDFs) at two distinct times and applying linear wave theory, they reported obliquely propagating $\mathcal{LZ}$ waves, and estimated their polarization and wavenumber $k\lambda_D\sim 10^{-3}$, smaller than expected for beam-driven waves. This wavevector corresponds to a wave with $f\lesssim f_p$, consistent with a $\mathcal Z$ mode.

Theoretical and numerical studies have shown that, even though linear mode conversion at constant frequency of beam-generated $\mathcal{LZ}$ wave turbulence proves more efficient for generating  $\mathcal Z$-mode waves in weakly magnetized plasmas with non negligible density turbulence (\cite{Krafft2025}, \cite{KrafftVolokitin2025}, \cite{Polanco2025b}), nonlinear decay of Langmuir/$\mathcal Z$-mode waves serves as an effective mechanism in solar wind regions where density turbulence is weak and the plasma is nearly homogeneous  (\cite{Krafft2024,Polanco2025a,Polanco2026a}). This work  aims to show that such decay process is responsible for the $\mathcal Z$-mode waves observed by Solar Orbiter on the 22 September 2022 at 13:52:28.

Through a detailed analysis of the corresponding data, we present the first observational evidence in the solar wind of the nonlinear three-wave decay of $\mathcal{LZ}$ waves  into electromagnetic $\mathcal{Z}$-mode radiation at $f_p$. Our findings are supported by multiple lines of evidence: waveform analysis revealing high cross-bicoherence levels, fulfillment of frequency and wavenumber resonance conditions, assessment of the decay threshold and turbulence parameter, examination of low-frequency wave dynamics, exclusion of alternative generation mechanisms, and consistency with our earlier theoretical predictions (\cite{Polanco2025a,Polanco2026a}). 

In addition, we present a second event detected by Solar Orbiter, which further confirms the same nonlinear mechanism at work. Our analysis integrates magnetic waveform observations with wave phase coherence analysis, two-dimensional theoretical modeling, and direct comparison to Particle-in-Cell (PIC) simulations which replicate the observed decay processes. Together, these elements provide, compared to previous analyses on $\mathcal{LZ}$ wave decay, an unprecedented level of evidence, robustly identifying the underlying physical mechanism.

Furthermore, the direct comparison of waveforms recorded in simulations and by Solar Orbiter enriches the discussion and provides deeper insights. It enables us to confirm the wave generation mechanism while inferring key properties of the density turbulence in the plasma region where the waves are observed, thereby clarifying its role in the wave dynamics. Finally, while linear mode conversion at constant frequency is typically more efficient than wave decay for $\mathcal{Z}$-mode radiation at $f_p$,  the latter can become dominant when wave packets reach very large amplitudes in extended density wells where they are trapped. The Solar Orbiter observations presented here align with this scenario.

\section{Instruments and  data acquisition}\label{Instruments}

The data analyzed below have been recorded by the Time Domain Sampler (TDS) of the Radio and Plasma Waves (RPW) instrument onboard Solar Orbiter (\cite{Maksimovic2020}).  The TDS (\cite{Soucek2021}) provides survey mode measurements of  electric and magnetic field components, with a sampling rate of $262$ kHz, offering highly resolved waveforms with typical durations of $60$ ms. The recorded data consist of two electric field components $E_y$ and $E_z$---obtained through the measurements of three voltages between coplanar antennas---and a magnetic field component $B_x$ perpendicular to $E_y$ and $E_z$. The axes  $x$, $y$, and $z$ correspond to the Satellite Reference Frame (SRF). During normal operation, the TDS captures a waveform snapshot each second. Onboard software then analyzes each of them, classifies it into distinct categories, and determines if it is valuable to be transferred to the Earth.

The two snapshots analyzed below, chosen as representative examples from a broader set of observations, were captured on 22 of September 2022 at 13:52:28 UT and 14:05:04 UT. A type III solar radio burst was being observed simultaneously to the spacecraft's encounter with a relativistic electron beam (\cite{Formanek2025}). In the SFR, the solar wind speed measured at 13:52:28 UT by the Solar Wind Analyzer (SWA) is $\mathbf V_{sw}\simeq (-325.1,-11,5)$ km/s,  while the onboard Magnetometer (MAG) provides the ambient magnetic field $\mathbf B_0=(-9.4,11.6,5)$ nT. Later, at 14:05:04 UT,  these values have slightly changed, as $\mathbf V_{sw}\simeq (-326,-2,0)$ km/s and $\mathbf B_0=(-8.7,9.5,7.4)$ nT. Measurements of the plasma frequency were done by the Thermal Noise Analyzer (TNR) at 13:52 UT and 14:14 UT, providing $f_p\simeq47.2\textnormal{ kHz}$ and $f_p\simeq58.5\textnormal{ kHz}$, respectively. Then, the magnetization ratio can be estimated as $f_c/f_p\simeq0.01$ for both events.  
On the other hand, the beam energy  and the plasma electron temperature measured at 13:38 UT are $E_b\simeq 23.9$ keV and $T_e \simeq11.7$ eV (\cite{Formanek2025}), which leads  to $v_b \simeq 0.21c\simeq 43.7v_T$ and $v_T \simeq 0.0048c$. A second measurement is available at 14:05 UT,  when the electron beam is decelerated, providing  $T_e=11.9$ eV.

\section{Evidence of nonlinear wave decay leading to $\mathcal{Z}$-mode waves}

\subsection{Description of the event at 13:52:28 UT}\label{Description event}

 \begin{figure*}[htb!]
    \centering
    \includegraphics[width=0.9\linewidth]{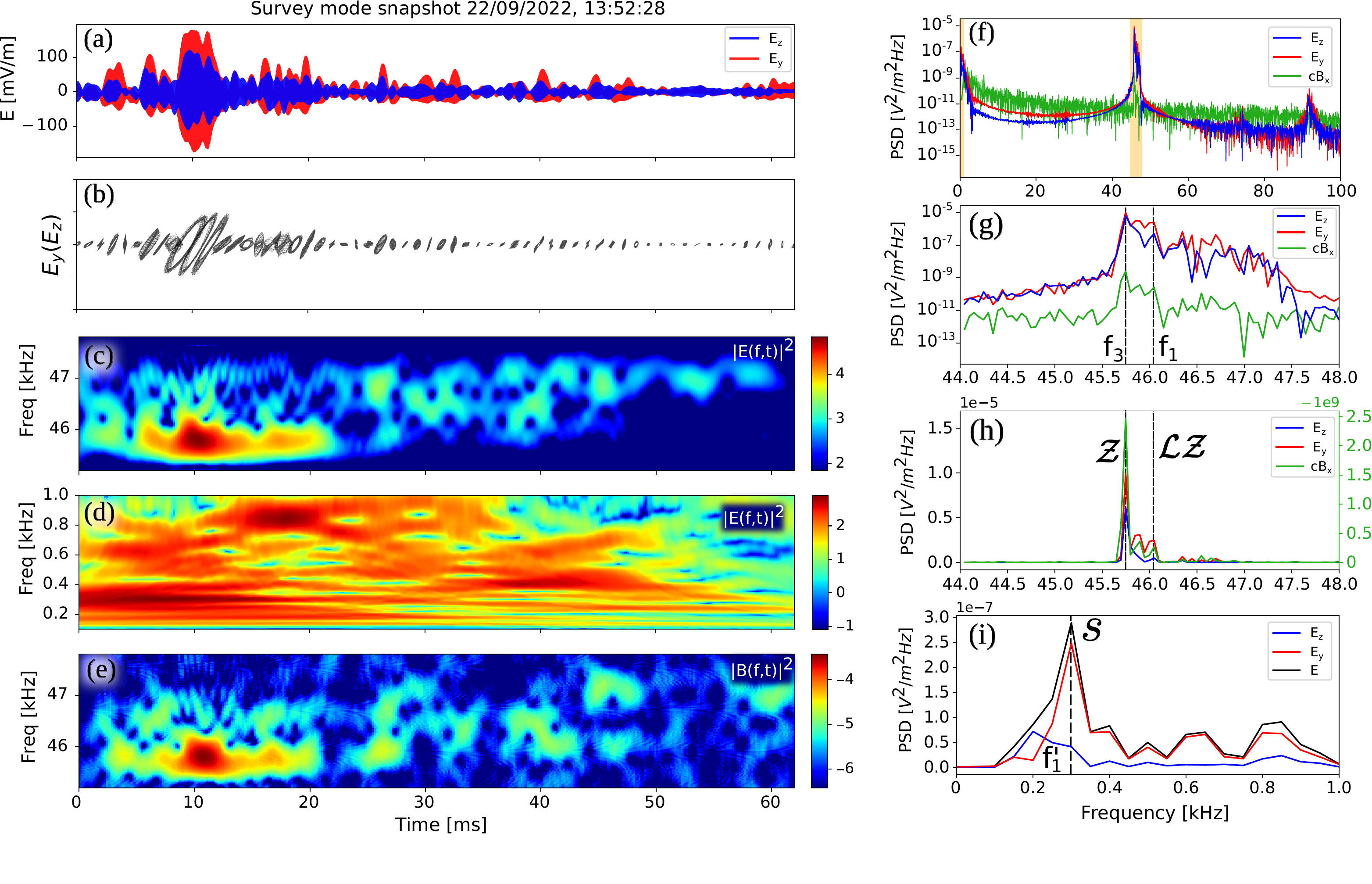}
    \caption{Snapshot captured in the survey mode by the Solar Orbiter  instrument RPW, on 22 September 2022, at 13:52:28 UT. (a) Waveform of the two electric field components $E_z$ (blue) and $E_y$ (red),  in the SRF frame. (b) Hodograms $E_y(E_z)$ calculated within equidistant time windows of $0.7$ ms. (c-d) Wavelet spectrograms of the electric energy $|E(f,t)|^2=|E_y(f,t)|^2+|E_z(f,t)|^2$ in the high- and low-frequency ranges, i.e. 45 $\leq f\leq48$ kHz and $f \leq 1$ kHz, respectively. (e) Wavelet spectrogram  of the magnetic energy $|B(f,t)|^2=|B_x(f,t)|^2$, in the same frequency range as (c). (c-e) : the color bars are in logarithmic scales. (f-i) Power spectra of the  field components $E_z$ (blue), $E_y$ (red)  and $cB_x$ (green),  respectively. Zoomed-in views are presented,  in logarithmics and linear scales,  in the ranges $44\leq f\leq48$ kHz (g-h) and  $f\leq1$ kHz (i); the black curve in (i) represents the spectrum of $|E|^2$. The dashed vertical lines  correspond to peaks at  $f_{3}\simeq45.74$ kHz, and $f_{1}\simeq46.04$ kHz (g-h), and at $f_{1}'\simeq0.3$ kHz (i). Spectra in (f-i) are calculated in the time interval 2 ms $\leq t\leq22$ ms.}
    \label{fig:snapshot111}
\end{figure*}

Figure \ref{fig:snapshot111} presents the event recorded by  Solar Orbiter on 22 September 2022 at 13:52:28 UT, during the detection of a type III burst-related electron beam. The waveforms of the electric field components $E_y(t)$ and $E_z(t)$ are shown in Figure \ref{fig:snapshot111}(a). A large-amplitude wave packet reaching $150$ mV/m is observed around $t\simeq 10$ ms. At the same time, the hodograms in panel (b) clearly demonstrate elliptic polarization, excluding purely electrostatic waves. 

Meanwhile, the high-frequency electric energy spectrogram shows wave packets excited around $f\simeq45.8$ kHz. 
At the same time, the low-frequency spectrogram  exhibits ion acoustic waves mainly excited near $f\sim 0.3$ kHz, below the ion plasma frequency $f_{pi}\simeq 1$ kHz (Figure \ref{fig:snapshot111}(d)).
Finally, the high-frequency magnetic energy spectrogram (Figure \ref{fig:snapshot111}(e)) shows a distinct emission at the same time and frequencies as the electric energy, consistent with the expected behavior of electromagnetic $\mathcal Z$-mode waves.

 Figure \ref{fig:snapshot111}(f) displays the power spectra of the fields $E_z$, $E_y$ and $B_x$ across a wide frequency range, all showing strong excitation near $f\sim46$ kHz. Zoomed-in views  in Figures \ref{fig:snapshot111}(g-h) provide precise measurements of the most intense spectral peaks at $f_1\simeq46.04$ kHz, $f_2\simeq45.87$ kHz, and $f_3\simeq45.74$ kHz,  excited with  $(E/cB)^2\sim 10^4$. The magnetic energy spectrum mirrors these peaks, while the low-frequency electric energy spectrum features a broad, prominent peak at $f'_1\simeq0.3$ kHz—without a magnetic counterpart above the noise level. 
 
The frequency peaks fall within the expected frequency range for beam-generated $\mathcal{LZ}$ waves.  Indeed, the $\mathcal{LZ}$ wave Doppler-shifted frequencies are given by $f_{\mathcal{LZ}}= f^{sw}_{\mathcal{LZ}}(\mathbf{k})+\mathbf k\cdot\mathbf V_{sw} \simeq f_p(1+3k^2\lambda_D^2/2+(V_{sw}/v_T) k\lambda_D\cos\theta)$, where $ f^{sw}_\mathcal{LZ}(\mathbf{k})$ are the frequencies in the solar wind frame;  $\theta$ is the angle between the $\mathcal{LZ}$ wave vector  $\mathbf k$ and the  solar wind velocity $\mathbf{V}_{sw}$  ($V_{sw}\simeq 330 $ km/s $\simeq0.23v_T$). Note that the magnetic corrections to this dispersion are second-order in $f_c$ and  thus negligible here, given that $f_c/f_p\simeq0.01$. The frequency range of beam-driven $\mathcal{LZ}$ waves can then be estimated  for any $\theta$ as $45.6\leq f_{\mathcal{LZ}}\leq 46.06$ kHz, which aligns closely with the frequency band in which the large-amplitude wave packet is observed (Figure \ref{fig:snapshot111}(f-h)). We have considered that  $k\lambda_D\simeq k_b\lambda_D\simeq0.023$, where $k_b$ is the characteristic wavenumber of beam-driven $\mathcal{LZ}$ waves,  and have defined the plasma frequency as $f_p\simeq f_3\simeq 45.74$  kHz, as explained below.

\subsection{Evidence for $\mathcal{LZ}$ wave decay} 

We show below that the emissions observed in Figure \ref{fig:snapshot111}, that were identified by \cite{Formanek2025} as oblique propagating $\mathcal{LZ}$ waves with small wavenumbers $k\lambda_D\sim10^{-3}$, arise from nonlinear wave decay, as supported by the arguments presented below. 
\begin{enumerate}[label=\textit{\roman*}.\vspace{0.3mm}]
    \item Frequency and wavenumber resonance conditions between three time-synchronized waves are fulfilled.
    \item The observed wave dynamics aligns with theoretical predictions of $\mathcal{LZ}$ wave decay in weakly magnetized plasmas.
    \item Quantitative estimates of the wave turbulence parameter and the decay threshold support the occurrence of this process.
    \item The two other alternative mechanisms can be excluded.
    \item Cross-bicoherence diagnostics confirm strong phase coherence between the interacting waves. 
\end{enumerate}

Further sections provide additional conclusive evidence, including the observation of a second event with similar characteristics and wave decay, along with 2D PIC simulations that remarkably reproduce the observed waveforms and validate the underlying mechanism.

\subsubsection{Frequency and wavevectors' three-wave resonance conditions}\label{Evidences: resonance}

The waves emitted at frequencies $f_1$, $f_3$, and  $f_1'$ in Figures \ref{fig:snapshot111}(g-i) satisfy the  frequency resonance condition $f_1\simeq f_3+f'_1$ with a very good accuracy, as $f_1\simeq46.04\textnormal{ kHz}$ and $f_3+f'_1\simeq (45.74 + 0.3)\textnormal{ kHz}=46.04\textnormal{ kHz}$. Because this relationship connects Doppler-shifted frequencies, it also holds for their wavevectors  in the solar wind frame. Furthermore, spectrograms  indicate that the three corresponding wave packets appear simultaneously around $t\sim10$ ms (Figures \ref{fig:snapshot111}(c-e)). These findings  suggest that the wave emitted at the lowest frequency $f_{\mathcal Z}=f_3$, which carries the highest magnetic energy, may correspond to a $\mathcal{Z}$-mode wave generated through the decay of a mother $\mathcal{LZ}$ wave excited at $f_{\mathcal{LZ}}=f_1$, together with ion acoustic waves $\mathcal{S}$ appearing at $f_{\mathcal{S}} =f'_1$, through the  channel  $\mathcal{LZ}\longrightarrow\mathcal{Z}+\mathcal{S}$. If this hypothesis holds—and further evidence supports it—then the dispersive properties of  $\mathcal{Z}$-mode waves and the kinematics of the decay imply that these waves must be produced just below $f_p$ but very close to it (\cite{Polanco2025a}). Based on this assumption, the plasma frequency at the exact moment of the waveform recording can be estimated as $f_p\simeq45.74$ kHz. 

The intensity of the $\mathcal{LZ}$ mother waves  is observed to be approximately five times lower than that  of the daughter $\mathcal Z$-mode waves. This difference can be explained by two mechanisms that may occur simultaneously. First, the decay  results in the transfer of energy from $\mathcal{LZ}$ waves to both $\mathcal{Z}$-mode and  $\mathcal S$ waves. Second, in a randomly inhomogeneous plasma, such as the solar wind, a tail of accelerated beam electrons is formed as a result of $\mathcal{LZ}$ wave scattering on density fluctuations $\delta n$, even when their average level $\Delta N=\langle(\delta n/n_0)^2 \rangle^{1/2}$ is very small. These electrons can reabsorb the energy of $\mathcal{LZ}$ waves (\cite{Krafft2013, KrafftSavoini2023}). Such a suprathermal tail is visible  in the VDF measured by Solar Orbiter during the event (see Figure 2 in \cite{Formanek2025}). Note that daughter waves have been observed with higher intensity than their mother wave (e.g. \cite{Kellogg2013}).

 \subsubsection{Dynamics of $\mathcal{LZ}$ wave decay in a weakly magnetized plasma}
 The three wave packets observed at frequencies $f_1$, $f_3$, and  $f_1'$  are consistent with a single-step  $\mathcal{LZ}$ wave decay cascade. Indeed, the wavenumber shift between the mother and daughter high-frequency waves can be approximated by $k_0 \lambda_D = 2c_s/3v_T$, where $c_s$ is the ion acoustic velocity (e.g., \cite{Layden2013},  \cite{KrafftSavoini2024}). For typical values of $0.3 \lesssim T_i / T_e\lesssim 1$ at $0.5$ au (\cite{Dakeyo2022}), we get that $ k_0\lambda_D\sim0.02$. 
Since $k_0 / k_b \simeq  1$ ($k_b$ is the wavevector of beam driven $\mathcal{LZ}$ waves), only a single decay cascade is expected. Then, the mother $\mathcal{LZ}$ waves transfer their energy directly to $\mathcal{Z}$-mode waves, without redistributing it among other $\mathcal{LZ}$ waves. However, while these estimates are approximate, they do not exclude the possibility of two decay cascades. Observations indicate that the $\mathcal Z$-mode waves are a terminal decay product. If they were not, their high amplitude would lead to further decay—a process that is not observed here.

In the  one-dimensional  approximation, decay cascades of $\mathcal{LZ}$ wave terminate at  $k\lambda_D\simeq k_*\lambda_D=(v_T/c)(1+f_p/f_c)^{-1/2}$ (\cite{CairnsLayden2018}),  i.e. at $k\lambda_D\simeq  4.7\cdot 10^{-4}$ for the plasma conditions of the observed event. In two-dimensional geometry, PIC simulations reveal the formation of  a boundary curve $ k_*(\theta)$  around $k\sim0$, on which the energy of decaying waves is ultimately deposited. It can be determined by calculating the iso-contour $f_{\mathcal{Z}}(k_\parallel,k_\perp)=f_0$ of the $\mathcal{Z}$-mode wave dispersion relation, where $f_0$ is the  frequency at which the group velocity of $\mathcal{Z}$-mode waves sharply increases, marking  the termination of the decay process (\cite{Polanco2025a}). Applying the plasma parameters of the observed event, we find that  $4.7\cdot 10^{-4}\lesssim k_*(\theta)\lambda_D\lesssim1.02\cdot10^{-3}$ for any $\theta$. By combining electron beam measurements with a linear wave dispersion analysis, \cite{Formanek2025} identified the emission at $f\simeq45.74$ kHz as an oblique $\mathcal{LZ}$ wave with small wavenumber $k\lambda_D\simeq10^{-3}$. This result aligns closely with our theoretical predictions. Rather than coincidental, this consistency provides further evidence supporting the decay mechanism while helping to rule out alternative mechanisms (section \ref{Evidences: exclusion EMD LMC}).

Finally, for the decay process to proceed,  the group velocities of $\mathcal{S}$ and $\mathcal{LZ}$ waves must be of similar orden of magnitude. Under typical solar wind conditions at 0.5 au, we find that $ c_s/v_T=((m_e/m_i)(1+3T_i/T_e))^{1/2}\simeq 0.033$, while $v_g^{\mathcal{LZ}}/v_T\simeq 3k_b\lambda_D\simeq0.069$, enabling effective energy exchange between the waves  during their propagation.

\subsubsection{$\mathcal{LZ}$ wave turbulence intensity}

The observed wave packet displays an exceptionally high intensity, likely driven by the simultaneously detected non-relaxed electron beam. Despite its large amplitude, there is no indication of strong turbulence phenomena. To verify that wave decay is indeed the mechanism generating the observed $\mathcal{Z}$-mode waves, we calculate the turbulence parameter of $\mathcal{LZ}$ waves as $W_{\mathcal{LZ}}=\varepsilon_0|E|_{\mathcal{LZ}}^2/4 n_0k_BT_e\simeq 3.5\cdot 10^{-4}$, where  $|E|_{\mathcal{LZ}}\simeq86$ mV/m is the peak electric field amplitude of  $\mathcal{LZ}$ waves,  estimated from the electric field waveform filtered near $f_p$. In the $(k\lambda_D,W)$ parameter space, as defined by Zakharov’s classification  (\cite{Zakharov1985}, \cite{Robinson1997}), the point $(k_b\lambda_D,W_{\mathcal{LZ}})$ lies within Region I, where  wave decay instability dominates —excluding other processes such as modulational instabilities. 

Moreover, for the decay process $\mathcal{LZ}\longrightarrow\mathcal{Z}+\mathcal{S}$ to proceed,  the energy carried by the $\mathcal{LZ}$ waves must exceed a specific threshold. This condition  can be expressed as ${\varepsilon_0E_{\mathcal{LZ}}^2}/{n_0k_BT_e}>4({\gamma_{\mathcal S'}}{\gamma_{\mathcal{LZ'}}})/({f_{\mathcal S'}}{f_{\mathcal{LZ'}}})$ (e.g., \cite{Shukla1983}), where $f_{\mathcal S'}$  and $f_{\mathcal{LZ'}}$ denote the frequencies of $\mathcal S'$ and $\mathcal{LZ'}$ waves, and $\gamma_{\mathcal S'}$  and ${\gamma_{\mathcal{LZ'}}}$ represent their damping rates, respectively. Calculations indicate that the required electric field threshold—on the order of a few mV/m (\cite{Lin1986}, \cite{GrahamCairns2013b}) —is significantly exceeded. 

\subsubsection{Exclusion of alternative mechanisms} \label{Evidences: exclusion EMD LMC}

Two other mechanisms might \textit{a  priori} seem capable of explaining the observed electromagnetic emission at  $f_p$. The first is the linear mode conversion process (LMC) at constant frequency  which can generate large-amplitude electromagnetic waves in randomly inhomogeneous plasmas (\cite{KrafftSavoini2022a}, \cite{Krafft2025}, \cite{KrafftVolokitin2025}). However, multiple considerations indicate that LMC is not the dominant mechanism at play here. 
First, the mechanism would fail to reproduce the sharp $\mathcal{Z}$-mode peak observed near $f_p$ (Figure \ref{fig:snapshot111}(h)). Instead, scattering on density fluctuations would result in a broader, smoother peak extending toward lower frequencies down to $\mathcal{Z}$-mode cutoff frequency. Second, since $f_c/f_p \ll 1$,  LMC would generate $\mathcal{X}$-mode waves with energy comparable to that of $\mathcal{Z}$-mode waves (\cite{Krafft2025}),  as well as  $\mathcal{O}$-mode waves, which are not observed. Third,  if LMC were the primary mechanism, the bicoherence diagnostics and the resonance conditions would lack the clarity and precision found above  in section \ref{Evidences: Bicoherence}, due to their random modifications through wave scattering on density fluctuations. 

An alternative mechanism to consider is electromagnetic decay, where $\mathcal{LZ}$ waves generate $\mathcal{O}$-mode and ion acoustic $\mathcal{S}$ waves through the channel $\mathcal{LZ}\longrightarrow\mathcal O+\mathcal S$. While this process could excite low-frequency waves within the recorded frequency ranges, PIC simulations—using parameters closely matching the observations—indicate that the ratio $(E/cB)^2$ reaches $\sim10^4$ for $\mathcal{Z}$-mode waves (in agreement with observations, see section \ref{Description event}), whereas  for $\mathcal{O}$-mode waves, it is at least an order of magnitude lower (not shown here). By other means, \cite{Formanek2025} also report  that  $(E/cB)^2\simeq 10^{4}$ for the observed peak. Consequently, the energy peak at $f_3\simeq45.74$ kHz cannot be attributed to $\mathcal{O}$-mode waves. 

Finally, other nonlinear wave–wave interactions involving low-frequency modes—beyond ion-acoustic waves—can be ruled out. First, no magnetic energy is observed at the corresponding  frequencies. Second, in weakly magnetized plasmas such as those studied here ($f_c/f_p \ll 1$), PIC simulations have shown that the low-frequency waves driven by the beam-generated $\mathcal{LZ}$  wave turbulence are ion acoustic modes (\cite{Polanco2025a}).

\subsection{Cross-bicoherence diagnostics}\label{Evidences: Bicoherence}

To confirm the occurrence of three-wave nonlinear decay, let us  look for phase coherence between interacting waves and thus calculate the cross-bicoherence level $b_c$ defined in the frequency  plane $(f_1,f_2)$  as 
\begin{equation}
        b_c(f_1,f_2)=\frac{|\langle X^n(f_1)Y^n(f_2)Z^{n*}(f_1+f_2)\rangle_n|}{\sqrt{\langle|X^n(f_1)|^2|Y^n(f_2)|^2\rangle_n\langle|Z^n(f_1+f_2)|^2\rangle_n}},
        \label{bico}
\end{equation}
where $X(t)$, $Y(t)$ and $Z(t)$ are three wave fields;  $X^n(f)$ is the Fourier transform of $X(t)$, calculated in the time window $n\Delta t\leq t\leq n\Delta t+\Delta T$ of duration $\Delta T$; $\Delta t$ is the time interval between two successive windows. The averaging $\langle \;\;\rangle_n$ is performed over different sliding windows indexed by $n$. The cross-bicoherence level $b_c$ varies between $0$ and $1$, equals $1$ when the fields $X$, $Y$, and $Z$ present a perfect phase coherence at $f_1$, $f_2$ and $f_3=f_1+f_2$, respectively, and equals $0$ in the absence of phase coherence.

\begin{figure}[htb!]
    \centering
    \includegraphics[width=\linewidth]{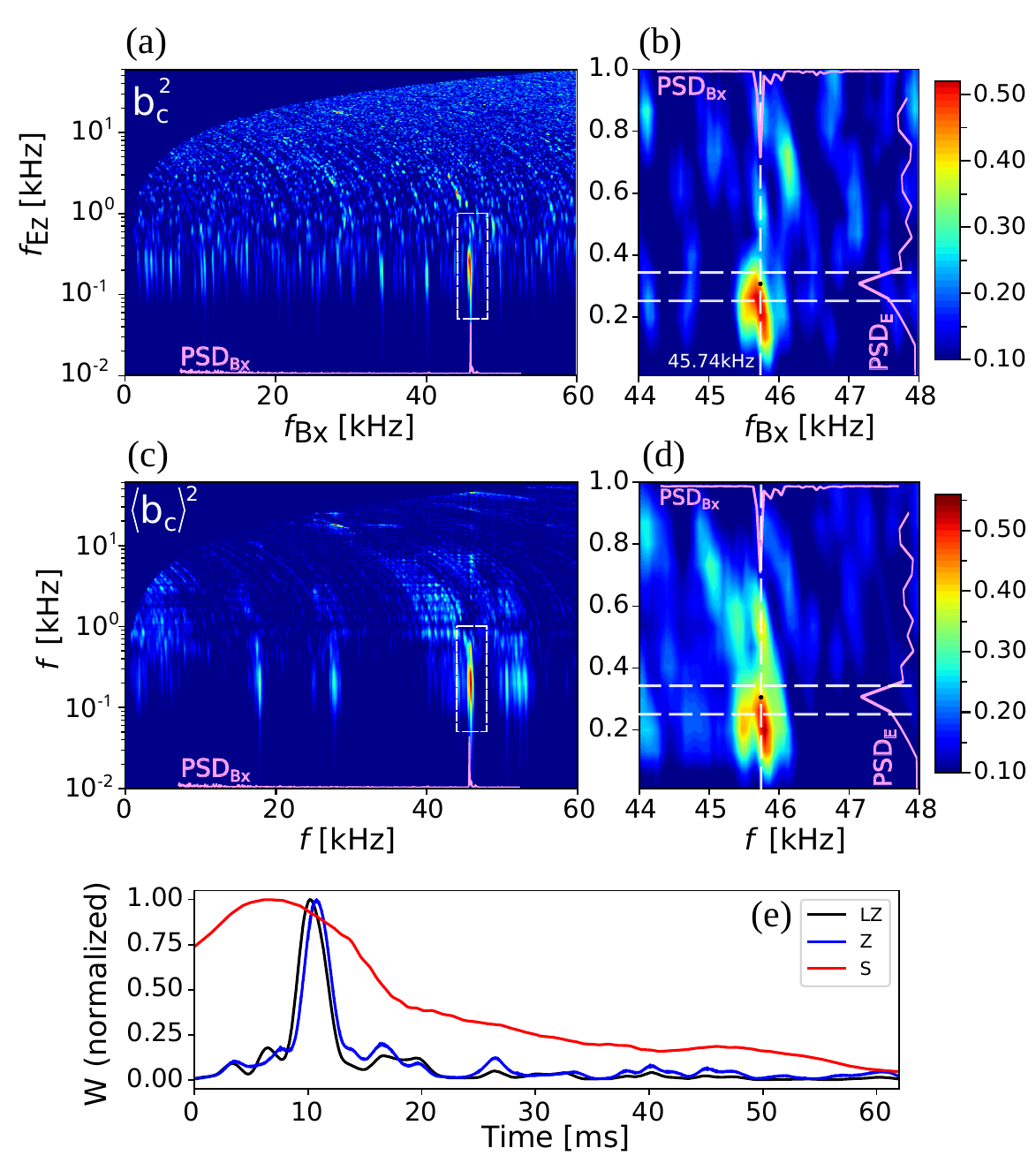}
    \caption{ Determination of phase coherence between waves using cross-bicoherence. (a) Square of the cross-bicoherence $b_c^2(f_{B_x},f_{E_z})$ computed with the field triad $(B_x,E_z,B_x)$ over the time interval $0 \leq t \leq 62$ ms, with $\Delta T=15$ ms and  $\Delta t=1$ ms (see equation \ref{bico} and text), in the frequency ranges $f_{B_x}\leq 60$ kHz and  $f_{E_z}  \leq 60$ kHz. (b) Zoom-in  of (a) in the map region 44 kHz $ \leq f_{B_x} \leq 48$ kHz and $f_{E_z}\leq 1$ kHz, with  $b_c\simeq0.7$ at $(f_{B_x},f_{E_z})=(f_\mathcal{Z},f_\mathcal{S})\simeq(45.74,0.3)$ kHz (black point). (c) Square of the cross-bicoherence averaged on 18 different triads with $E_z$ or $E_y$ in the second position, $\langle b_c(f_1,f_2)\rangle^2$, in the same frequency range as (a). (d) Zoom-in of (c) in the same frequency range as in (b), with  $b_c\simeq0.7$ at $(f_{B_x},f_{E_z})\simeq(45.74,0.3)$ kHz (black point). (b,d) : The white vertical lines indicate the local  plasma frequency. The magenta curves represented on the borders of the panels refer to the high- and low-frequency spectra used to calculate $b_c$ (see Figures \ref{fig:snapshot111}(h,i)). (e) Variations with time of the $\mathcal{LZ}$, $\mathcal{Z}$ and $\mathcal{S}$ wave energies obtained by integration of the wavelet spectrograms of $E$, $B$ and $E$ fields, respectively, in the ranges $45.2\textnormal{ kHz}\leq f\leq 47.8\textnormal{ kHz}$ for $\mathcal{LZ}$ and $\mathcal{Z}$-mode waves, and  $f\leq0.4\textnormal{ kHz}$ for $\mathcal S$ waves. }
    \label{fig:snapshot_111_bico}
\end{figure}

Figure \ref{fig:snapshot_111_bico} shows the squared cross-bicoherence level, $b_c^2$, computed for the field triad ($B_x,E_z,B_x$). The results are displayed in the  $(f_{B_x},f_{E_z})$ plane across a broad frequency range (panel (a)), as well as in focused views of the narrower frequency domains corresponding to the measured peaks (Figure \ref{fig:snapshot_111_bico}(b)). The bicoherence reaches the high level $b_c\simeq0.7$ at $(f_{B_x},f_{E_z})=(f_\mathcal{Z},f_\mathcal{S})\simeq(45.74,0.3)$ kHz,  which is a signature of  the decay  $\mathcal{LZ}\longrightarrow\mathcal Z+\mathcal S$, corresponding to the resonance condition $f_{\mathcal{LZ}} \simeq f_{\mathcal{Z}}+f_{\mathcal{S}}$, with $f_{\mathcal{LZ}} \simeq 46.04$ kHz.

\begin{figure*}[htb!]
    \centering
    \includegraphics[width=0.9\linewidth]{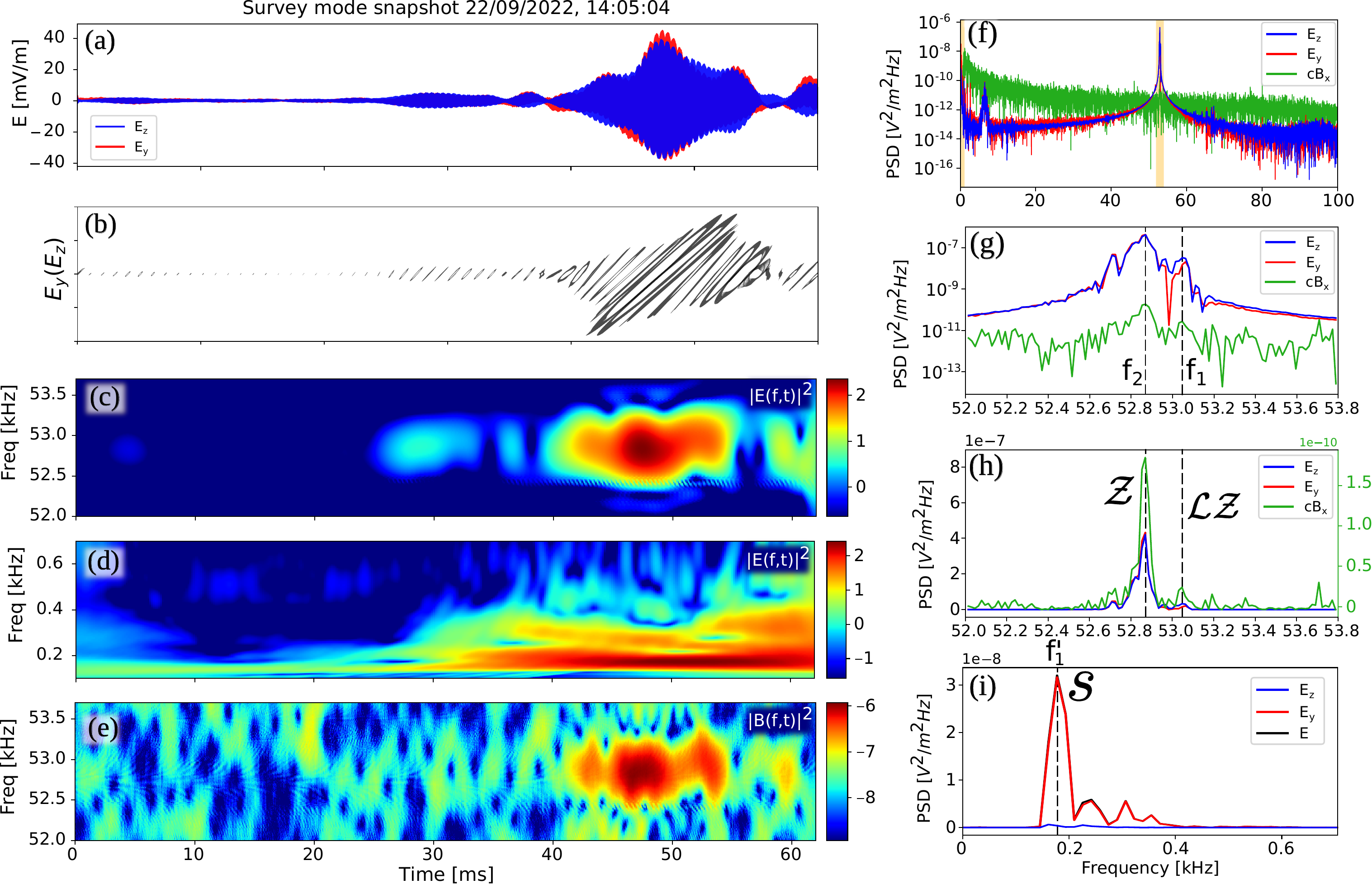}
    \caption{Snapshot captured in the survey mode by RPW
       on 22 September 2022, at 14:05:04 UT. (a)  Waveform of the two electric field components $E_z$ (blue) and $E_y$ (red),  in the SRF frame. Hodograms $E_y(E_z)$ calculated within equidistant time windows of $0.7$ ms. (c-d)  Wavelet spectrograms of the electric energy $|E(f,t)|^2$ in the high- and low-frequency ranges, i.e. 45 kHz $\leq f\leq48$ kHz and $f \leq 0.7$ kHz, respectively. (e) Wavelet spectrogram  of the magnetic energy $|B(f,t)|^2$, in the same frequency range as (c). (c-e) : the color bars are in logarithmic scales. (f-i) Power spectra of the  field components $E_z$ (blue), $E_y$ (red)  and $cB_x$ (green),  respectively. Zoomed-in view are presented,  in linear scales,  in the ranges 45 kHz $\leq f\leq48$ kHz (g-h) and at  $f\leq1.7$ kHz (i). (h-i) : linear scales;  the black curve represents the spectrum of $|E|^2$. The vertical lines  correspond to peaks at $f_{3}\simeq52.71$ kHz and $f_{1}\simeq53.05$ kHz  (g-h), and at $f_{1}'\simeq0.18$ kHz (i). Spectra in (f-i) are calculated in the time interval $0\leq t\leq62$ ms.}
    \label{fig:snapshot113}
\end{figure*}

While the full frequency range includes small isolated regions with elevated cross-bicoherence levels (Figure \ref{fig:snapshot_111_bico}(a)), these features correspond to waves with energies close to the noise floor (see Figure \ref{fig:snapshot111}(f)), and and should therefore be dismissed  as non-physical artifacts. Although sheath effects can sometimes produce high $b_c$ (\cite{Graham2014}), our analysis focuses mostly on triads involving the magnetic component $B_x$, which is unaffected by such effects. In Figures \ref{fig:snapshot_111_bico}(a,b),  the triad $(B_x,E_z,B_x)$ reveals elevated  $b_c$ levels exclusively in the region $(f_\mathcal{Z},f_\mathcal{S})$.  This precise localization effectively rules out sheath effects as the source of the observed phase coherence. Moreover, the detected waves not only display phase coherence but also satisfy resonance conditions and exhibit precise temporal synchronization—features that render spacecraft-related artifacts an improbable explanation for these observations. 

To validate the robustness of our analysis, we compute the cross-bicoherence for all possible triads involving either $E_y$ or $E_z$  in the second position (totaling $18$ combinations) and average over the results (Figures \ref{fig:snapshot_111_bico}(c,d)). The cross-bicoherence level remains consistently high ($b_c\simeq 0.7$), at the same frequencies as in panel (a), confirming that the nonlinear wave interaction identified in panels (a,b) persists across all triads.
While Figure \ref{fig:snapshot_111_bico}(c) shows lower bicoherence peaks, these do not correspond to any physical signal, as evidenced in Figure \ref{fig:snapshot111}(f). Since these spurious artifacts are absent in some triads (see panel (a)), their bicoherence level reduces through averaging. 

Finally, Figure \ref{fig:snapshot_111_bico}(e)  illustrates the time evolution of the energies carried by the $\mathcal{LZ},\mathcal S$ and $\mathcal{Z}$-mode waves, derived by integrating the electric and magnetic spectrograms over their respective frequency bands (see caption). A simultaneous rise in the energy of all three modes is observed, with a slight temporal offset of up to 1 ms. The close synchronization of the energy peaks highlights the strong temporal correlation between the three modes. 

The presented arguments strongly supports wave decay as the dominant mechanism generating the observed $\mathcal Z$-mode waves. However,  it may not act alone.  Indeed, in plasmas with average levels $\Delta N$ of random density fluctuations of a few percent, the linear transformations of $\mathcal{LZ}$ waves on inhomogeneities can stimulate nonlinear wave-wave interactions (\cite{Polanco2026a}). Moreover, plasma regions with gentle density fluctuations may also facilitate decay (\cite{KrafftSavoini2024}, \cite{Krafft2024}). By comparing observations with numerical simulations, we show that they are consistent with a scenario where a  $\mathcal{LZ}$ wave packet is trapped in a broad,  flat-bottomed density well, where a decay process  can unfold.

\section{Evidence of decay to $\mathcal{Z}$-mode waves during a second snapshot}

In a snapshot recorded on 22 September 2022 at 14:05:04 UT---around 12 mn after that presented in  section \ref{Description event}---clear signatures of $\mathcal Z$-mode waves generated through $\mathcal{LZ} $ wave decay are again detected.  Figure \ref{fig:snapshot113} presents the corresponding waveforms and power spectra, presented in the same way as in  Figure \ref{fig:snapshot111}. At  14:05:04 UT,  the beam is significant relaxed  (\cite{Formanek2025}). Panel (a)  highlights the emergence of a wave packet at  $t\sim50$ ms, with an amplitude reaching $\sim40$ mV/m—approximately one-fifth of the amplitude observed at 13:52:28 UT. This reduction likely stems from beam relaxation, which lowers the overall intensity of $\mathcal{LZ} $ wave turbulence.  The waveform exhibits clear amplitude modulation, suggesting the simultaneous presence of different high-frequency waves. 
The hodograms reveal a pattern close to that observed at 13:52:28 UT, suggesting that the waves share the same polarization properties. At $t\simeq45$ ms,  both the electric and magnetic energy spectrograms display pronounced peaks at frequency $f\simeq$ 53 kHz, consistent with the  $\mathcal Z$-mode wave radiation (Figures \ref{fig:snapshot113}(c,e)). Simultaneously, low-frequency emissions peak at $f\sim 0.2$ kHz (Figure \ref{fig:snapshot113}(d)).

Figures \ref{fig:snapshot113}(f-g) show the electric and magnetic energy spectra over a broad frequency range. A zoomed-in  view reveals three distinct high-frequency peaks at $f_{1} = 53.05$ kHz, $f_{2} = 52.86$ kHz, and $f_{3} = 52.71$ kHz. Meanwhile, the low-frequency spectrum exhibits a single peak at $f_{1}' = 0.18$ kHz (Figures \ref{fig:snapshot113}(i)). The triad $(f_{2}, f_{1}, f_{1}')$ satisfies readily the resonance condition $f_{1} \simeq f_{2} + f_{1}'$  with high accuracy, supporting the decay process $\mathcal{LZ}\longrightarrow\mathcal Z+\mathcal S$.  Notably, the peak at  $f_{2} = 52.86$ kHz reaches a pronounced maximum in the magnetic spectrum, confirming its identification as a  $\mathcal Z$-mode wave. The ratio $(E/cB)^2$  decreases from $10^4$ at $53.05$ kHz down to $5\cdot 10^3$ at $52.86$ kHz, which is consistent with the decay of  mother $\mathcal{LZ}$ waves into electromagnetic $\mathcal Z$-mode waves near the boundary curve $k_*(\theta)$.  On the other hand, the peak at $f_{3} = 52.71$ kHz does not satisfy any resonance relation with the low-frequency wave packet and is therefore unrelated to the wave coupling. Likewise, as done previously, one easily checks that the threshold is overcome and that $\varepsilon_0E^2/4n_0k_BT\simeq 2\cdot 10^{-5}$ (with $k\lambda_D>0.023$, as the beam velocity is lower than in the previous event), so that no strong turbulence process is expected (\cite{Robinson1997}). Note that in Figure \ref{fig:snapshot113}(g,h), the electric and magnetic spectra extend slightly down $f_{} \simeq 52.80$ kHz, just below the dominant peak. This subtle feature could indicate the presence of  weak scattering and LMC  concurrently with the primary nonlinear process.

\begin{figure}[htb]
    \centering
    \includegraphics[width=\linewidth]{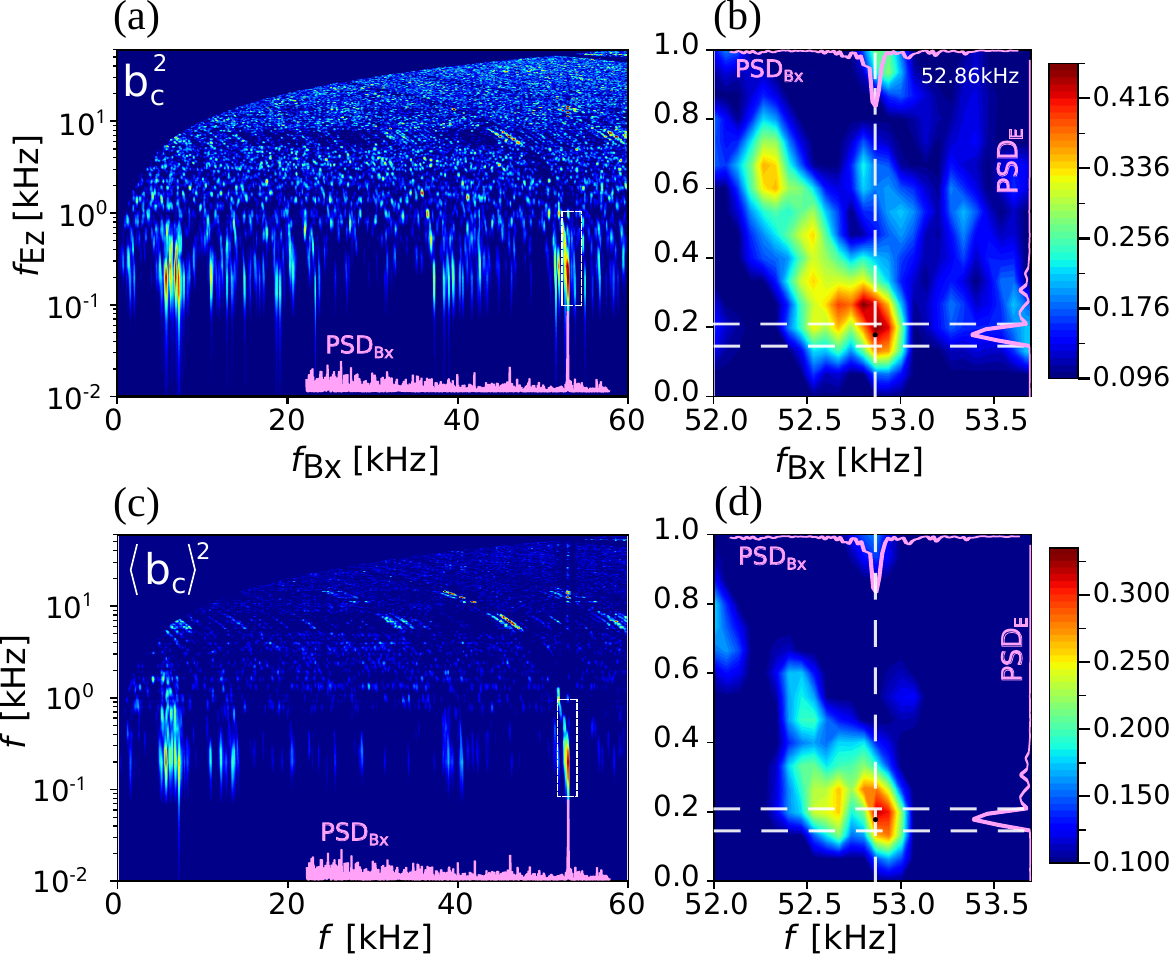}
    \caption{Determination of the phase coherence between waves using cross-bicoherence. (a) Square of the cross-bicoherence $b_c^2(f_{B_x},f_{E_y})$ computed with the field triad $(B_x,E_y,E_y)$ over the time interval $0 \leq t \leq 62$ ms, with $\Delta T=15$ ms and  $\Delta t=1$ ms  (see equation \ref{bico} and text), in the frequency range $ f_{B_x},f_{E_y} \leq 60$ kHz. (b) Zoom-in of (a) in the domain $52\textnormal{ kHz} \leq f_{B_x} \leq 53.7$ kHz and $f_{E_y} \leq 1$ kHz, with  $b_c\simeq0.65$ at $(f_{B_x},f_{E_y})=(f_\mathcal{Z},f_\mathcal{S})\simeq(52.86, 0.18)$ kHz (black point). (c) Square of the cross-bicoherence averaged over the 18 triads with $E_y$ or $E_z$ in the second position, $\langle b_c(f_1,f_2)\rangle^2$, in the same  frequency range as (a). (d) Zoom-in of (c),  in the same frequency domain as in (b), with   $b_c\simeq0.55$ at $(f_{B_x},f_{E})=(f_\mathcal{Z},f_\mathcal{S})$ kHz (black point). (b,d) : The white vertical line indicates the local  plasma frequency. The pink curves represented on the borders of the panel are the high- and the low-frequency spectra used to calculate $b_c$ (see Figures  \ref{fig:snapshot113}(h,i)).}
    \label{fig:snapshot_113_bico}
\end{figure}

The nonlinear decay of  $\mathcal{LZ}$ waves at $f_{\mathcal{LZ}}\simeq53.05$ kHz is confirmed in Figure \ref{fig:snapshot_113_bico}(b) by the high cross-bicoherence level $b_c \simeq 0.65$ found at  $(f_{\mathcal Z},f_{\mathcal S})=(f_{E_y}, f_{B_x}) \simeq (52.86, 0.18)$ kHz. This corresponds precisely to the peaks of the high- and low-frequency spectra. When averaged over the same 18 triads as described in Section \ref{Evidences: Bicoherence}, the cross-bicoherence decreases slightly to $b_c\simeq0.55$, while effectively suppressing noise and artifacts at other frequencies (panel (d)).  Note that wave energy of unidentified origin  is observed at $f \sim 8$ kHz, accompanied by elevated $b_c$ values. However, as these waves are off-resonance with respect to the waves under study, they fall outside the scope of the present work. 

\section{Comparison with numerical simulations}

\begin{figure*}[htb]
    \centering
    \includegraphics[width=\linewidth]{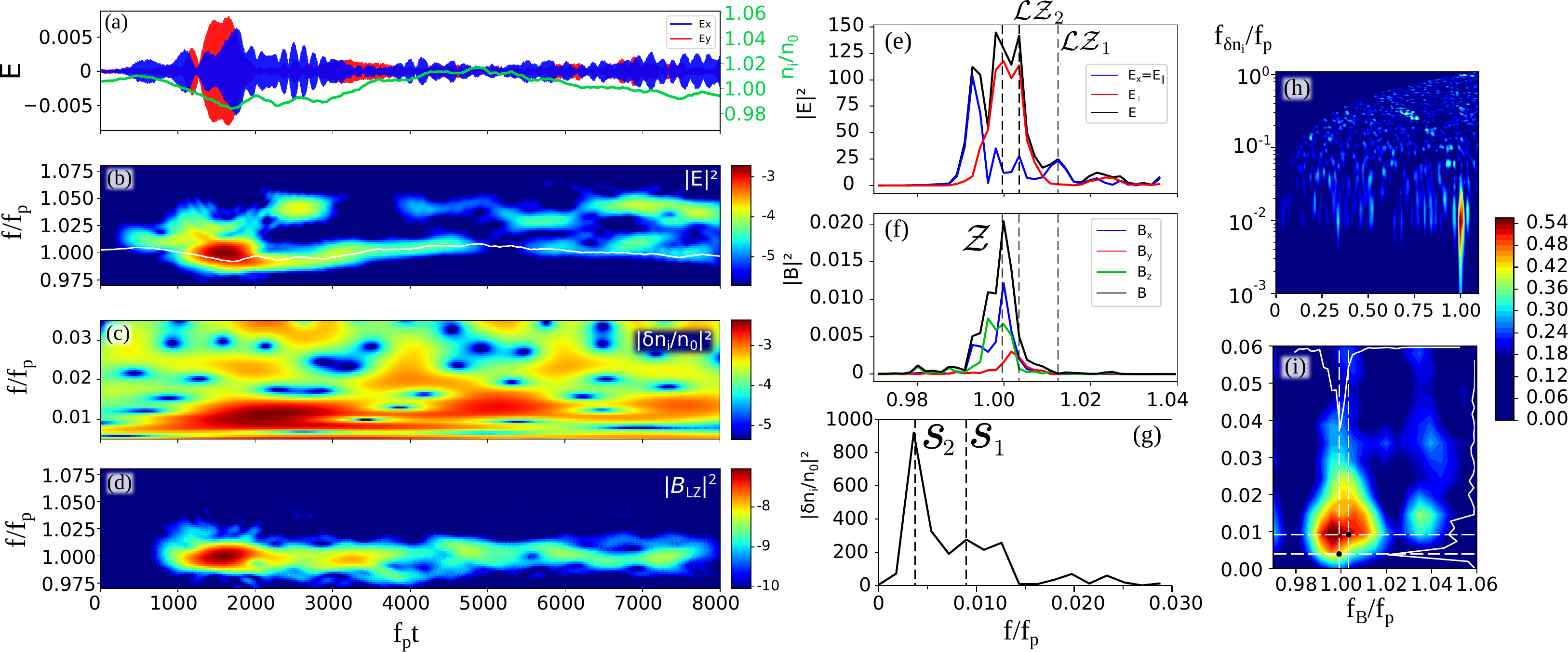}
    \caption{Waveform recorded by a virtual satellite moving  into a 2D PIC simulation plane with the velocity $|v_{s}|=0.1v_T$, in the direction opposite to $\mathbf{B}_0$. Plasma and beam parameters are  $\Delta N=\langle (\delta n/n_0)^2\rangle^{1/2}=0.025$,  $f_c/f_p=0.02$, and $v_b=0.25c$.  (a) Waveforms of the parallel $E_x=E_\parallel$ (blue) and perpendicular $E_{\perp}\simeq E_y$ (red) electric fields ($E_z$ is negligibly small). The green line represents the  normalized ion density $n_i/n_0$. (b)  Spectrograms of the electric energy $|E(f,t)|^2=|E_x(f,t)|^2+|E_y(f,t)|^2+|E_z(f,t)|^2$ in the frequency range $0.97\leq f/f_p\leq1.08$; the white line represents the local plasma frequency $f_{pl}$ normalized to the average plasma frequency $f_p$. (c)  Spectrogram of  $|\delta n_i(f,t)/n_0|^2$ in the range $0.005\leq f/f_p\leq0.035$. (d) Spectrogram of $|B(f,t)|^2$ in the same frequency range as (b). (e) Power spectra of the electric field energy  $|E(f)|^2$ (black), in the range $0.97\leq f/f_p\leq1.06$; parallel and perpendicular energies are shown in blue and red, respectively. (f) Power spectra of the magnetic field energy  $|B(f)|^2$ (black), in the same frequency range as (e); the three field components are indicated in color.  (e-f) : spectra are calculated in the time interval $500\leq f_pt\leq4000$. (g) Low-frequency power spectrum $|\delta n_i(f)/n_0|^2$  in the range $f/f_p\leq0.035$. (e-g) : vertical dashed lines indicate the frequencies $f_3\simeq0.999f_p$, $f_2\simeq1.004f_p$ and $f_1\simeq1.012f_p$ (e,f) as well as $f_3'\simeq0.004f_p$, $f_2'\simeq0.009f_p$ and $f_1'\simeq0.0125f_p$ (g). (h-i) Square of the cross-bicoherence averaged on the  4 triads  $(B_x, \delta n_i,E_x)$, $(B_y,\delta n_i,E_x)$, $(B_x,\delta n_i,E_y)$ and $(B_y,\delta n_i,E_y)$, in the ranges $f/f_p\leq1.1$ (h), as well as $0.97\leq f_1/f_p\leq1.06$ and $ f_2/f_p\leq0.06$ (i). Sliding windows of duration $\Delta T=1300f_p^{-1}$, separated by $\Delta t=800f_p^{-1}$, are employed to calculate $b_c$ over the time range $f_pt\leq5000$. }
    \label{fig:simulation}
\end{figure*}

Comparisons with 2D/3V PIC simulations can shed light on the above studies by offering a more complete view and detailed understanding of the mechanisms at play. Figure \ref{fig:simulation}(a) displays  a waveform recorded by a virtual satellite moving in the simulation plane with the velocity $|v_{s}| = 0.1v_T$, in the direction antiparallel to the ambient magnetic field  $\mathbf B_0$. This is consistent with the direction of the solar wind flow toward the outer Solar System. Main beam and plasma parameters match as close as possible the conditions expected near 0.5 au or measured by Solar Orbiter on September 22, 2022, i.e. $f_c/f_p=0.02$,  $v_b=0.25c$, and $\Delta N={\langle(\delta n/n_0)^2\rangle}^{1/2}=0.025$  (see for more details \cite{KrafftSavoini2021,  KrafftSavoini2022a, Krafft2024, Krafft2025}). Dimensionless variables are used below. In particular, the time and frequencies are normalized by $f_p$, the ion density $n_i$ by the average plasma density $n_0$. Fields and energies are presented in arbitrary units. As the beam velocity, the ambient magnetic field is directed along the $x$-axis; the simulation plane is $ (x,y)$.

The electric field waveform features a distinct wave packet centered at  $t \simeq 1500f_p^{-1}$, exhibiting significant amplitudes in both the parallel and perpendicular components with respect to  $\mathbf B_0$ (Figure  \ref{fig:simulation}(a)). It is observed as the virtual satellite crosses a shallow density depletion, where the ratio of the ion density $n_i$ to the average plasma density $n_0$ reaches $n_i/n_0\simeq0.985$ over the time interval   $ [500, 4500]f_p^{-1}$.  The spectrogram of the electric energy demonstrates that the wave packet is composed of $\mathcal{LZ}$ waves with frequencies distributed around the local plasma frequency $f_{pl}$ (white line in Figure \ref{fig:simulation}(b)), due to weak wave scattering on $\delta n$. At later times, weaker high-frequency energy emerges slightly above $f_{pl}$, indicating the presence of large-$k$ (Doppler-shifted) $\mathcal{LZ}$ waves,  which are characterized by minimal magnetic signatures.

The low-frequency energy spectrogram in Figure~\ref{fig:simulation}(c) reveals pronounced  ion acoustic wave emissions  around $ 0.01f_p$.  Meanwhile, Figure \ref{fig:simulation}(d), which presents the spectrogram of the magnetic energy $|B_{\mathcal{LZ}}(f,t)|^2=|B_x(f,t)|^2+|B_y(f,t)|^2$ carried by the  $\mathcal{LZ}$  waves  (\cite{Polanco2025a}), displays large-amplitude emissions around $f_p$. Those temporally coincide with the intense low- and high-frequency electric wave activity.
At this point, a striking similarity emerges between the waveforms and spectrograms of Figures  \ref{fig:snapshot111} and \ref{fig:simulation}. As a consequence, the analysis  is now enriched by new parameters, including the local plasma frequency, the six field components, the ion and electron densities, as well as the full set of beam-plasma conditions. 

Figures \ref{fig:simulation}(e-g) present the high-frequency electric and magnetic spectra, along with the low-frequency one, which is obtained here from the ion density perturbations $\delta n_i=n_i-n_0$ directly, and not from the low-frequency electric fields, as in the above sections. Figure \ref{fig:simulation}(e) highlights four  ${\mathcal{LZ}}$ wave energy peaks, at $f_4\simeq0.992f_p$, $f_3\simeq0.999f_p$, $f_2\simeq1.004f_p$, and $f_1\simeq 1.012f_p$. This spectral complexity reflects the underlying wave turbulence trapped within the density well ($ 500\lesssim f_pt \lesssim4500
$). In this gently inhomogeneous region, several wave packets coexist simultaneously, though not all participate in the nonlinear decay.
Furthermore, they correspond to significant magnetic signatures with $(E/cB)^2\sim 10^{4}$ (Figure \ref{fig:simulation}(f)), except at $f_4 $ and $f_1$ where $(E/cB)^2\simeq 2\cdot 10^5$  and  $(E/cB)^2\simeq3\cdot 10^6$, respectively.
The peak observed at $f_3=f_{\mathcal Z}\simeq0.999f_p$ is attributed to ${\mathcal{Z}}$-mode waves generated through the nonlinear decay of ${\mathcal{LZ}}$ waves, as discussed below.  In contrast, the peak at $f_4\simeq0.992f_p$ corresponds to  ${\mathcal{LZ}}$ waves, likely resulting from scattering on density fluctuations,  Doppler-shift effects, or a combination of both. Since the peak is off-resonance with other waves, it is not analyzed further. Finally, the spectrum $|\delta n_i(f)/n_0|^2$ of the ion density perturbations  reveals ion acoustic waves with energy maxima at $f_3'=0.004f_p$,  $f_2'=0.009f_p$, and $f_1'=0.0125f_p$  (Figure \ref{fig:simulation}(g)).

Two three-wave frequency resonance conditions can be identified : $f_{\mathcal{ LZ }_1} \simeq 1.012f_p \simeq f_{\mathcal{LZ}_2}+f_{\mathcal S_2} \simeq 1.004f_p+0.009f_p$, and $f_{\mathcal{LZ}_2}\simeq 1.004f_p \simeq f_{\mathcal Z} +f_{\mathcal S_3}\simeq0.999f_p+0.004f_p$. 
In the first scenario,  the waves $\mathcal{LZ}_1$ decay into daughter waves through the channel $\mathcal{LZ}_1\longrightarrow\mathcal {LZ}_2+\mathcal S_2$, while the second one describes a subsequent decay cascade, $\mathcal{LZ}_2\longrightarrow\mathcal Z+\mathcal S_3$, which deposits high-frequency wave energy at $f_{\mathcal {Z} }\simeq 0.999f_p$. It is worth noting that the $\mathcal{LZ}_1$ waves are not directly driven by the beam. Instead, they likely arise from an alternative wave process. Their relatively small wavevectors enable them to decay into  $\mathcal Z$-mode waves through just two successive decay cascades. 

The decay processes are corroborated by cross-bicoherence diagnostics, as illustrated  in Figures  \ref{fig:simulation}(h,i), which reveal strong phase coherence within both the wave triads ($\mathcal{LZ}_1,\mathcal {LZ}_2,\mathcal S_2$) and ($\mathcal{LZ}_2,\mathcal {Z},\mathcal S_3$), respectively.  The cross-bicoherence levels,  obtained by averaging over four different triads (see the caption) reach $\langle b_c\rangle\simeq0.65$ at $(f_{\mathcal Z},f_{\mathcal S_3})=(0.999,0.004)f_p$,  and $\langle b_c\rangle\simeq0.73$ at $(f_{\mathcal{LZ}_2},f_{\mathcal S_2})=(1.004,0.009)f_p$ (panel (i)). Additionally, Figure  \ref{fig:simulation}(h)  shows that the significant cross-bicoherence levels are strictly confined to the small frequency ranges of Figure  \ref{fig:simulation}(i). 

Many of the features observed in this simulation closely mirror those captured in the Solar Orbiter snapshots. First, the localization of the large-amplitude wave packets, as shown in Figures \ref{fig:snapshot111} and \ref{fig:snapshot113}, can be attributed to wave trapping within an extensive and nearly flat-bottomed density well. Notably, in nearly homogeneous plasmas with minimal density turbulence, $\mathcal{LZ}$ wave energy does not exhibit such localization (\cite{Polanco2025b, Polanco2026a}). Second, similar to the simulation, the wave packets recorded by Solar Orbiter include frequency peaks that do not satisfy resonance conditions and  lack the significant phase coherence typically associated with nonlinear three-wave interactions. This behavior is characteristic of wave turbulence, where spectral complexity is further amplified by the scattering of $\mathcal{LZ}$ wave packets on the density fluctuations.

In addition, the higher-frequency waves observed in Figure \ref{fig:snapshot111} at $t\gtrsim40$ ms and $f\gtrsim46$ kHz exhibit a similar behavior to that shown in Figure \ref{fig:simulation}, where the variations of the local plasma frequency induces $\mathcal{LZ}$ waves' frequency shifts. This  suggests that the wave packets are propagating in a  solar wind plasma with gentle density fluctuations. Furthermore, as discussed in Section \ref{Evidences: resonance}, the predicted frequency range for $\mathcal{LZ}$ waves is very narrow, ruling out Doppler shifts as the origin of these higher-frequency waves.

\section{Conclusion}

During a type III radio burst, the Solar Orbiter spacecraft encountered an electron beam and captured high-resolution electric and magnetic waveforms.  Based on these observations, this study presents the first evidence of the nonlinear decay of  beam-driven $\mathcal{LZ}$ waves into electromagnetic $\mathcal Z$-mode radiation at $f_p$ in the solar wind. The three-wave interaction process is demonstrated through several key findings:  the resonance conditions for frequency and wavenumber are satisfied, strong phase coherence exists between the waves,  the interacting modes coincide temporally,  and the results align closely with theoretical predictions—including the decay threshold, the turbulence parameter of $\mathcal{LZ}$ waves,  and the two-dimensional decay kinematics and dynamics. 
Additionally, a second waveform demonstrating the same decay process is presented. Particle-in-cell simulations, conducted under close beam-plasma conditions, successfully replicate the observations. Notably, they suggest that the wave packets observed by Solar Orbiter may be trapped within extended, nearly flat-bottomed density wells, where the decay process is not overcome by wave scattering on random density fluctuations and subsequent mode conversion effects.

These results stem from recent key developments: first, the progress in \textit{in situ} magnetometry, which allowed for the unambiguous identification of $\mathcal Z$-mode waves;  second, the foundational theoretical work—validated through Particle-in-Cell (PIC) simulations—that enabled the identification of the process driving their generation; and third, the methodology that connects \textit{in situ} spacecraft measurements and large-scale, long-term 2D/3V PIC simulations through the virtual satellite technique. Modern missions such as Parker Solar Probe and  Solar Orbiter offer an ideal platform for applying and further developing this approach.

\section{Acknowledgements}

 This work was granted access to the HPC computing and storage resources under the allocation 2023-A0130510106 and 2024-A017051010 made by GENCI. This research was also financed in part by the French National Research Agency (ANR) under the project ANR-23-CE30-0049-01.  C.K. thanks the International Space Science Institute (ISSI) in Bern through ISSI International Team project No. 557, Beam-Plasma Interaction in the Solar Wind and the Generation of Type III Radio Bursts. C. K. thanks the Institut Universitaire de France (IUF). F.J.P.R. thanks T. Chust and D. Fontaine for offering their expertise in satellite data. 

Solar Orbiter is a mission of international cooperation between ESA and NASA, operated by ESA. The RPW instrument has been designed and funded by CNES, CNRS, the Paris Observatory, the Swedish National Space Agency, ESA-PRODEX and all the participating institutes.
 
 For open access purposes, a CC-BY license has been applied by the authors to this document and will be applied to any subsequent version up to the author's manuscript accepted for publication resulting from this submission.
   

\printbibliography

@ARTICLE{Bale2000,
       author = {{Bale}, S.~D. and {Larson}, D.~E. and {Lin}, R.~P. and {Kellogg}, P.~J. and {Goetz}, K. and {Monson}, S.~J.},
        title = "{On the beam speed and wavenumber of intense electron plasma waves near the foreshock edge}",
      journal = {\jgr},
         year = 2000,
        month = dec,
       volume = {105},
       number = {A12},
        pages = {27353-27368},
          doi = {10.1029/2000JA900042},
       adsurl = {https://ui.adsabs.harvard.edu/abs/2000JGR...10527353B}
}

@article{Bale1998,
 adsurl = {https://ui.adsabs.harvard.edu/abs/1998GeoRL..25....9B},
 author = {{Bale}, S.~D. and {Kellogg}, P.~J. and {Goetz}, K. and {Monson}, S.~J.},
 doi = {10.1029/97GL03493},
 journal = {\grl},
 month = {01},
 number = {1},
 pages = {9-12},
 title = {{Transverse z-mode waves in the terrestrial electron foreshock}},
 volume = {25},
 year = {1998}
}

@article{Bale2016,
 adsurl = {https://ui.adsabs.harvard.edu/abs/2016SSRv..204...49B},
 author = {{Bale}, S.~D. and {Goetz}, K. and {Harvey}, P.~R. and {Turin}, P. and {Bonnell}, J.~W. and {Dudok de Wit}, T. and {Ergun}, R.~E. and {MacDowall}, R.~J. and {Pulupa}, M. and {Andre}, M. and {Bolton}, M. and {Bougeret}, J.-L. and {Bowen}, T.~A. and {Burgess}, D. and {Cattell}, C.~A. and {Chandran}, B.~D.~G. and {Chaston}, C.~C. and {Chen}, C.~H.~K. and {Choi}, M.~K. and {Connerney}, J.~E. and {Cranmer}, S. and {Diaz-Aguado}, M. and {Donakowski}, W. and {Drake}, J.~F. and {Farrell}, W.~M. and {Fergeau}, P. and {Fermin}, J. and {Fischer}, J. and {Fox}, N. and {Glaser}, D. and {Goldstein}, M. and {Gordon}, D. and {Hanson}, E. and {Harris}, S.~E. and {Hayes}, L.~M. and {Hinze}, J.~J. and {Hollweg}, J.~V. and {Horbury}, T.~S. and {Howard}, R.~A. and {Hoxie}, V. and {Jannet}, G. and {Karlsson}, M. and {Kasper}, J.~C. and {Kellogg}, P.~J. and {Kien}, M. and {Klimchuk}, J.~A. and {Krasnoselskikh}, V.~V. and {Krucker}, S. and {Lynch}, J.~J. and {Maksimovic}, M. and {Malaspina}, D.~M. and {Marker}, S. and {Martin}, P. and {Martinez-Oliveros}, J. and {McCauley}, J. and {McComas}, D.~J. and {McDonald}, T. and {Meyer-Vernet}, N. and {Moncuquet}, M. and {Monson}, S.~J. and {Mozer}, F.~S. and {Murphy}, S.~D. and {Odom}, J. and {Oliverson}, R. and {Olson}, J. and {Parker}, E.~N. and {Pankow}, D. and {Phan}, T. and {Quataert}, E. and {Quinn}, T. and {Ruplin}, S.~W. and {Salem}, C. and {Seitz}, D. and {Sheppard}, D.~A. and {Siy}, A. and {Stevens}, K. and {Summers}, D. and {Szabo}, A. and {Timofeeva}, M. and {Vaivads}, A. and {Velli}, M. and {Yehle}, A. and {Werthimer}, D. and {Wygant}, J.~R.},
 doi = {10.1007/s11214-016-0244-5},
 journal = {\ssr},
 month = {December},
 number = {1-4},
 pages = {49-82},
 title = {{The FIELDS Instrument Suite for Solar Probe Plus. Measuring the Coronal Plasma and Magnetic Field, Plasma Waves and Turbulence, and Radio Signatures of Solar Transients}},
 volume = {204},
 year = {2016}
}

@article{CairnsLayden2018,
 author = {Cairns, Iver H. and Layden, A.},
 doi = {10.1063/1.5037300},
 issn = {1070-664X},
 journal = {\pop},
 month = {08},
 number = {8},
 pages = {082309},
 title = {{Kinematics of electrostatic 3-wave decay of generalized Langmuir waves in magnetized plasmas}},
 volume = {25},
 year = {2018}
}

@article{Dakeyo2022,
 archiveprefix = {arXiv},
 author = {{Dakeyo}, Jean-Baptiste and {Maksimovic}, Milan and {D{\'e}moulin}, Pascal and {Halekas}, Jasper and {Stevens}, Michael L.},
 doi = {10.3847/1538-4357/ac9b14},
 eid = {130},
 eprint = {2207.03898},
 journal = {\apj},
 month = {12},
 number = {2},
 pages = {130},
 primaryclass = {astro-ph.SR},
 title = {{Statistical Analysis of the Radial Evolution of the Solar Winds between 0.1 and 1 au and Their Semiempirical Isopoly Fluid Modeling}},
 volume = {940},
 year = {2022}
}

@article{Formanek2025,
 adsurl = {https://ui.adsabs.harvard.edu/abs/2025ApJ...985L..29F},
 author = {{Form{\'a}nek}, Tom{\'a}{\v{s}} and {Santol{\'\i}k}, Ond{\v{r}}ej and {Sou{\v{c}}ek}, Jan and {P{\'\i}{\v{s}}a}, David and {Zaslavsky}, Arnaud and {Kretzschmar}, Matthieu and {Maksimovic}, Milan and {Owen}, Christopher J. and {Nicolaou}, Georgios},
 doi = {10.3847/2041-8213/add687},
 eid = {L29},
 journal = {\apjl},
 month = {June},
 number = {2},
 pages = {L29},
 title = {{Polarization Analysis of Type III Langmuir/Z-mode Waves with Coherent Magnetic Component Observations by Solar Orbiter}},
 volume = {985},
 year = {2025}
}

@article{Fox2016,
 author = {N. J. Fox and M. C. Velli and S. D. Bale and R. Decker and A. Driesman and R. A. Howard and J. C. Kasper and J. Kinnison and M. Kusterer and D. Lario and M. K. Lockwood and D. J. McComas and N. E. Raouafi and A. Szabo},
 doi = {10.1007/s11214-015-0211-6},
 issn = {15729672},
 issue = {1-4},
 journal = {Space Sci. Rev.},
 month = {12},
 pages = {7-48},
 publisher = {Springer Netherlands},
 title = {The Solar Probe Plus Mission: Humanity’s First Visit to Our Star},
 volume = {204},
 year = {2016}
}

@article{Graham2014,
 author = {Graham, D. B. and Cairns, Iver H.},
 eprint = {https://agupubs.onlinelibrary.wiley.com/doi/pdf/10.1002/2013JA019425},
 journal = {\jgr},
 number = {4},
 pages = {2430-2457},
 title = {Dynamical evidence for nonlinear Langmuir wave processes in type III solar radio bursts},
 volume = {119},
 year = {2014}
}

@article{GrahamCairns2013b,
 author = {{Graham}, D.~B. and {Cairns}, Iver H.},
 doi = {10.1002/jgra.50402},
 journal = {\jgr},
 month = {07},
 number = {7},
 pages = {3968-3984},
 title = {{Electrostatic decay of Langmuir/z-mode waves in type III solar radio bursts}},
 volume = {118},
 year = {2013}
}

@article{Henri2009,
 author = {{Henri}, P. and {Briand}, C. and {Mangeney}, A. and {Bale}, S.~D. and {Califano}, F. and {Goetz}, K. and {Kaiser}, M.},
 doi = {10.1029/2008JA013738},
 eid = {A03103},
 journal = {\jgr},
 month = {03},
 number = {A3},
 pages = {A03103},
 title = {{Evidence for wave coupling in type III emissions}},
 volume = {114},
 year = {2009}
}

@ARTICLE{Jannet2021,
       author = {{Jannet}, G. and {Dudok de Wit}, T. and {Krasnoselskikh}, V. and {Kretzschmar}, M. and {Fergeau}, P. and {Bergerard-Timofeeva}, M. and {Agrapart}, C. and {Brochot}, J.-Y. and {Chalumeau}, G. and {Martin}, P. and {Revillet}, C. and {Bale}, S.~D. and {Maksimovic}, M. and {Bowen}, T.~A. and {Brysbaert}, C. and {Goetz}, K. and {Guilhem}, E. and {Harvey}, P.~R. and {Leray}, V. and {Lorf{\`e}vre}, E.},
        title = "{Measurement of Magnetic Field Fluctuations in the Parker Solar Probe and Solar Orbiter Missions}",
      journal = {\jgr},
         year = 2021,
        month = feb,
       volume = {126},
       number = {2},
          eid = {e28543},
        pages = {e28543},
          doi = {10.1029/2020JA028543},
       adsurl = {https://ui.adsabs.harvard.edu/abs/2021JGRA..12628543J}
}

@article{Kellogg2013,
 author = {{Kellogg}, P.~J. and {Goetz}, K. and {Monson}, S.~J. and {Opitz}, A.},
 doi = {10.1002/jgra.50443},
 journal = {\jgr},
 month = {08},
 number = {8},
 pages = {4766-4775},
 title = {{Observations of transverse Z mode and parametric decay in the solar wind}},
 volume = {118},
 year = {2013}
}

@article{Krafft2013,
 author = {{Krafft}, C. and {Volokitin}, A.~S. and {Krasnoselskikh}, V.~V.},
 doi = {10.1088/0004-637X/778/2/111},
 eid = {111},
 journal = {\apj},
 month = {December},
 number = {2},
 pages = {111},
 title = {{Interaction of Energetic Particles with Waves in Strongly Inhomogeneous Solar Wind Plasmas}},
 volume = {778},
 year = {2013}
}

@article{Krafft2024,
 adsurl = {https://ui.adsabs.harvard.edu/abs/2024ApJ...967L..20K},
 author = {{Krafft}, C. and {Savoini}, P. and {Polanco-Rodríguez}, F.~J.},
 doi = {10.3847/2041-8213/ad47b5},
 eid = {L20},
 journal = {\apjl},
 month = {06},
 number = {2},
 pages = {L20},
 title = {{Mechanisms of Fundamental Electromagnetic Wave Radiation in the Solar Wind}},
 volume = {967},
 year = {2024}
}

@ARTICLE{Krafft2025,
       author = {{Krafft}, C. and {Volokitin}, A.~S. and {Polanco-Rodr{\'\i}guez}, F.~J. and {Savoini}, P.},
        title = "{Radiation efficiency of electromagnetic wave modes from beam-generated solar radio sources}",
      journal = {\na},
         year = 2025,
        month = sep,
       volume = {9},
        pages = {1292-1299},
          doi = {10.1038/s41550-025-02619-2},
archivePrefix = {arXiv},
       eprint = {2506.16816},
 primaryClass = {astro-ph.SR},
       adsurl = {https://ui.adsabs.harvard.edu/abs/2025NatAs...9.1292K}
}

@article{KrafftSavoini2021,
 author = {{Krafft}, C. and {Savoini}, P.},
 doi = {10.3847/2041-8213/ac1795},
 eid = {L23},
 journal = {\apjl},
 month = {08},
 number = {2},
 pages = {L23},
 title = {{Second Harmonic Electromagnetic Emissions by an Electron Beam in Solar Wind Plasmas with Density Fluctuations}},
 volume = {917},
 year = {2021}
}

@article{KrafftSavoini2022a,
 author = {{Krafft}, C. and {Savoini}, P.},
 doi = {10.3847/2041-8213/ac46a7},
 eid = {L24},
 journal = {\apjl},
 month = {01},
 number = {2},
 pages = {L24},
 title = {{Fundamental Electromagnetic Emissions by a Weak Electron Beam in Solar Wind Plasmas with Density Fluctuations}},
 volume = {924},
 year = {2022}
}

@article{KrafftSavoini2023,
 author = {{Krafft}, C. and {Savoini}, P.},
 doi = {10.3847/1538-4357/acc1e4},
 eid = {24},
 journal = {\apj},
 month = {05},
 number = {1},
 pages = {24},
 title = {{Dynamics of Two-dimensional Type III Electron Beams in Randomly Inhomogeneous Solar Wind Plasmas}},
 volume = {949},
 year = {2023}
}

@article{KrafftSavoini2024,
 author = {{Krafft}, C. and {Savoini}, P.},
 doi = {10.3847/2041-8213/ad3449},
 eid = {L30},
 journal = {\apjl},
 month = {04},
 number = {2},
 pages = {L30},
 title = {{Electrostatic Wave Decay in the Randomly Inhomogeneous Solar Wind}},
 volume = {964},
 year = {2024}
}

@ARTICLE{KrafftVolokitin2025,
       author = {{Krafft}, C. and {Volokitin}, A.~S.},
        title = "{Linear Mode Conversion Theory of Radio Emission from Turbulent Solar Wind Plasmas}",
      journal = {\apjs},
         year = 2025,
        month = dec,
       volume = {281},
       number = {2},
          eid = {67},
        pages = {67},
          doi = {10.3847/1538-4365/ae14f2},
archivePrefix = {arXiv},
       eprint = {2507.13856},
 primaryClass = {physics.plasm-ph},
       adsurl = {https://ui.adsabs.harvard.edu/abs/2025ApJS..281...67K}
}

@article{Krauss-Varban1989,
 adsurl = {https://ui.adsabs.harvard.edu/abs/1989JGR....94.3527K},
 author = {{Krauss-Varban}, D.},
 doi = {10.1029/JA094iA04p03527},
 journal = {\jgr},
 month = {04},
 number = {A4},
 pages = {3527-3534},
 title = {{Beam instability of the Z mode in the solar wind}},
 volume = {94},
 year = {1989}
}

@article{Larosa2022,
 author = {A. Larosa and T. Dudok de Wit and V. Krasnoselskikh and S. D. Bale and O. Agapitov and J. Bonnell and C. Froment and K. Goetz and P. Harvey and J. Halekas and M. Kretzschmar and R. MacDowall and David M. Malaspina and M. Moncuquet and J. Niehof and M. Pulupa and C. Revillet},
 doi = {10.3847/1538-4357/ac4e85},
 journal = {\apj},
 month = {03},
 number = {1},
 pages = {95},
 publisher = {The American Astronomical Society},
 title = {Langmuir-Slow Extraordinary Mode Magnetic Signature Observations with Parker Solar Probe},
 volume = {927},
 year = {2022}
}

@article{Layden2013,
 adsurl = {https://ui.adsabs.harvard.edu/abs/2013PhRvL.110r5001L},
 author = {{Layden}, A. and {Cairns}, Iver H. and {Li}, B. and {Robinson}, P.~A.},
 doi = {10.1103/PhysRevLett.110.185001},
 eid = {185001},
 journal = {\prl},
 month = {05},
 number = {18},
 pages = {185001},
 title = {{Electrostatic Decay in a Weakly Magnetized Plasma}},
 volume = {110},
 year = {2013}
}

@article{Lin1986,
 author = {{Lin}, R.~P. and {Levedahl}, W.~K. and {Lotko}, W. and {Gurnett}, D.~A. and {Scarf}, F.~L.},
 doi = {10.1086/164563},
 journal = {\apj},
 month = {September},
 pages = {954},
 title = {{Evidence for Nonlinear Wave-Wave Interactions in Solar Type III Radio Bursts}},
 volume = {308},
 year = {1986}
}

@article{Maksimovic2020,
 author = {M. Maksimovic and S. D. Bale and T. Chust and Y. Khotyaintsev and V. Krasnoselskikh and M. Kretzschmar and D. Plettemeier and H. O. Rucker and J. Souček and M. Steller and Š. Štverák and P. Trávníček and A. Vaivads and S. Chaintreuil and M. Dekkali and O. Alexandrova and P.-A. Astier and G. Barbary and D. Bérard and X. Bonnin and K. Boughedada and B. Cecconi and F. Chapron and M. Chariet and C. Collin and Y. de Conchy and D. Dias and L. Guéguen and L. Lamy and V. Leray and S. Lion and L. R. Malac-Allain and L. Matteini and Q. N. Nguyen and F. Pantellini and J. Parisot and P. Plasson and S. Thijs and A. Vecchio and I. Fratter and E. Bellouard and E. Lorfèvre and P. Danto and S. Julien and E. Guilhem and C. Fiachetti and J. Sanisidro and C. Laffaye and F. Gonzalez and B. Pontet and N. Quéruel and G. Jannet and P. Fergeau and J.-Y. Brochot and G. Cassam-Chenai and T. Dudok de Wit and M. Timofeeva and T. Vincent and C. Agrapart and G. T. Delory and P. Turin and A. Jeandet and P. Leroy and J.-C. Pellion and V. Bouzid and B. Katra and R. Piberne and W. Recart and O. Santolík and I. Kolmašová and V. Krupař and O. Krupařová and D. Píša and L. Uhlíř and R. Lán and J. Baše and L. Ahlèn and M. André and L. Bylander and V. Cripps and C. Cully and A. Eriksson and S.-E. Jansson and E. P. G. Johansson and T. Karlsson and W. Puccio and J. Břínek and H. Öttacher and M. Panchenko and M. Berthomier and K. Goetz and P. Hellinger and T. S. Horbury and K. Issautier and E. Kontar and S. Krucker and O. Le Contel and P. Louarn and M. Martinović and C. J. Owen and A. Retino and J. Rodríguez-Pacheco and F. Sahraoui and R. F. Wimmer-Schweingruber and A. Zaslavsky and I. Zouganelis},
 doi = {10.1051/0004-6361/201936214},
 issn = {0004-6361},
 journal = {A\&A},
 month = {10},
 pages = {A12},
 title = {The Solar Orbiter Radio and Plasma Waves (RPW) instrument},
 volume = {642},
 year = {2020}
}

@article{Malaspina2011,
 author = {Malaspina, David M. and Cairns, Iver H. and Ergun, Robert E.},
 doi = {10.1029/2011GL047642},
 journal = {\grl},
 number = {13},
 title = {Dependence of Langmuir wave polarization on electron beam speed in type III solar radio bursts},
 volume = {38},
 pages = {L13101},
 year = {2011}
}

@article{Muller2020,
 archiveprefix = {arXiv},
 author = {{M{\"u}ller}, D. and {St. Cyr}, O.~C. and {Zouganelis}, I. and {Gilbert}, H.~R. and {Marsden}, R. and {Nieves-Chinchilla}, T. and {Antonucci}, E. and {Auch{\`e}re}, F. and {Berghmans}, D. and {Horbury}, T.~S. and {Howard}, R.~A. and {Krucker}, S. and {Maksimovic}, M. and {Owen}, C.~J. and {Rochus}, P. and {Rodriguez-Pacheco}, J. and {Romoli}, M. and {Solanki}, S.~K. and {Bruno}, R. and {Carlsson}, M. and {Fludra}, A. and {Harra}, L. and {Hassler}, D.~M. and {Livi}, S. and {Louarn}, P. and {Peter}, H. and {Sch{\"u}hle}, U. and {Teriaca}, L. and {del Toro Iniesta}, J.~C. and {Wimmer-Schweingruber}, R.~F. and {Marsch}, E. and {Velli}, M. and {De Groof}, A. and {Walsh}, A. and {Williams}, D.},
 doi = {10.1051/0004-6361/202038467},
 eid = {A1},
 eprint = {2009.00861},
 journal = {\aap},
 month = {10},
 pages = {A1},
 primaryclass = {astro-ph.SR},
 title = {{The Solar Orbiter mission. Science overview}},
 volume = {642},
 year = {2020}
}

@article{Polanco2025a,
 author = {{Polanco-Rodr{\'\i}guez}, F.J. and {Krafft}, C. and {Savoini}, P.},
 doi = {10.3847/2041-8213/adba64},
 journal = {\apjl},
 month = {mar},
 number = {1},
 pages = {L24},
 publisher = {The American Astronomical Society},
 title = {Decay of Turbulent Upper-hybrid Waves in Weakly Magnetized Solar Wind Plasmas},
 url = {https://dx.doi.org/10.3847/2041-8213/adba64},
 volume = {982},
 year = {2025}
}

@ARTICLE{Polanco2025b,
       author = {{Polanco-Rodr{\'\i}guez}, F.J. and {Krafft}, C. and {Savoini}, P.},
        title = "{Polarization Ratios of Turbulent Langmuir/Z-mode Waves Generated by Electron Beams in Magnetized Solar Wind Plasmas}",
      journal = {\apjl},
         year = 2025,
        month = aug,
       volume = {989},
       number = {2},
          eid = {L38},
        pages = {L38},
          doi = {10.3847/2041-8213/adf4ce},
       adsurl = {https://ui.adsabs.harvard.edu/abs/2025ApJ...989L..38P}
}

@ARTICLE{Polanco2026a,
       author = {{Polanco-Rodr{\'\i}guez}, F.J. and {Krafft}, C. and {Savoini}, P.},
        title = "{Analysis of wave processes using beam-driven Langmuir/$\mathcal{Z}$-mode waveforms generated in Particle-In-Cell simulations}",
      journal = {arXiv e-prints},
         year = 2026,
        month = jan,
          eid = {arXiv:2601.09368},
        pages = {arXiv:2601.09368},
          doi = {10.48550/arXiv.2601.09368},
archivePrefix = {arXiv},
       eprint = {2601.09368},
 primaryClass = {physics.plasm-ph},
       adsurl = {https://ui.adsabs.harvard.edu/abs/2026arXiv260109368P}
}

@article{Robinson1997,
 author = {Robinson, P.~A.},
 journal = {Rev. Mod. Phys.},
 pages = {507--573},
 year = {1997},
 title = {Nonlinear wave collapse and strong turbulence},
 volume = {69}
}

@article{Shukla1983,
 author = {{Shukla}, P.~K. and {Yu}, M.~Y. and {Mohan}, M. and {Varma}, R.~K. and {Spatschek}, K.~H.},
 doi = {10.1103/PhysRevA.27.552},
 journal = {\pra},
 month = {01},
 number = {1},
 pages = {552-554},
 title = {{Electromagnetic wave generation in a beam-plasma system}},
 volume = {27},
 year = {1983}
}

@article{Soucek2021,
 adsurl = {https://ui.adsabs.harvard.edu/abs/2021A&A...656A..26S},
 author = {{Soucek}, J. and {P{\'\i}{\v{s}}a}, D. and {Kolmasova}, I. and {Uhlir}, L. and {Lan}, R. and {Santol{\'\i}k}, O. and {Krupar}, V. and {Kruparova}, O. and {Ba{\v{s}}e}, J. and {Maksimovic}, M. and {Bale}, S.~D. and {Chust}, T. and {Khotyaintsev}, Yu. V. and {Krasnoselskikh}, V. and {Kretzschmar}, M. and {Lorf{\`e}vre}, E. and {Plettemeier}, D. and {Steller}, M. and {{\v{S}}tver{\'a}k}, {\v{S}}. and {Vaivads}, A. and {Vecchio}, A. and {B{\'e}rard}, D. and {Bonnin}, X.},
 doi = {10.1051/0004-6361/202140948},
 eid = {A26},
 journal = {\aap},
 month = {12},
 pages = {A26},
 title = {{Solar Orbiter Radio and Plasma Waves - Time Domain Sampler: In-flight performance and first results}},
 volume = {656},
 year = {2021}
}

@article{Zakharov1985,
 adsurl = {https://ui.adsabs.harvard.edu/abs/1985PhR...129..285Z},
 author = {{Zakharov}, V.~E. and {Musher}, S.~L. and {Rubenchik}, A.~M.},
 doi = {10.1016/0370-1573(85)90040-7},
 journal = {\physrep},
 month = {December},
 number = {5},
 pages = {285-366},
 title = {{Hamiltonian approach to the description of non-linear plasma phenomena}},
 volume = {129},
 year = {1985}
}

\end{document}